\documentclass[aps,prd,reprint,superscriptaddress,nofootinbib,floatfix]{revtex4-2}
\usepackage{graphicx}% Include figure files
\usepackage{dcolumn}% Align table columns on decimal point
\usepackage{bm}% bold math
\usepackage{amsmath}
\usepackage{booktabs}
\usepackage{array}
\usepackage{placeins}
\usepackage{xcolor}
\begin{document}

\title{On the Contribution of Local Sources to the Galactic Cosmic-Ray Spectrum: \\
An Exact Series Solution for Two-Zone Diffusion}

\author{Zi-Hang Liu}
\affiliation{School of Astronomy and Space Science, Nanjing University, Nanjing 210023, China}
\affiliation{Key Laboratory of Modern Astronomy and Astrophysics (Nanjing University), Ministry of Education, Nanjing 210023, China}

\author{Yiwei Bao}
\email{sjtu0538015@sjtu.edu.cn}
\affiliation{Tsung-Dao Lee Institute, Shanghai Jiao Tong University, Shanghai 201210, China}
\affiliation{School of Physics and Astronomy, Shanghai Jiao Tong University, Shanghai 200240, China}

\author{Ruo-Yu Liu}
\email{ryliu@nju.edu.cn}
\affiliation{School of Astronomy and Space Science, Nanjing University, Nanjing 210023, China}
\affiliation{Key Laboratory of Modern Astronomy and Astrophysics (Nanjing University), Ministry of Education, Nanjing 210023, China}
\affiliation{Tianfu Cosmic Ray Research Center, Chengdu 610000, Sichuan, China}

\date{\today}

\begin{abstract}
Measurements of cosmic-ray proton and helium spectra below the knee show deviations from simple power laws, including multi-TeV structures.
A possible explanation is that one or a few nearby sources contribute an additional component to the local spectrum.
However, previous study shows that a dominant local contribution is statistically unlikely under a homogeneous diffusion model. In this
work, we investigate how this probability changes if cosmic rays experience inefficient transport near
their sources, motivated by observations of extended gamma-ray emission around Galactic accelerators.
We derive a series Green's function that enables fast calculation of the particle distribution in this scenario, making Monte Carlo calculations for Galactic source populations feasible.
The inner slow-diffusion region delays escape and redistributes the arriving particles in time and energy.
In Monte Carlo realizations, the probability that the strongest local source becomes comparable to the background at $10\,\rm{TeV}$ increases from about $0.4\%$ in homogeneous diffusion to $1.7$--$2.2\%$ in the two-zone models.
Thus inhibited near-source transport weakens, but does not remove, the statistical difficulty.
We then examine cataloged nearby candidate supernova remnants and show that a $10\,\rm{TeV}$ feature can be reproduced only with additional assumptions, especially a harder local injection spectrum and a favorable diffusion coefficient.
The predicted contribution of a given source changes strongly among different particle transport model.
Therefore, the local source interpretations are plausible but highly model dependent, and require independent constraints on source injection history, particle transport mechanisms, and local interstellar turbulence.
\end{abstract}

\maketitle

\section{Introduction}
\label{sec:introduction}

The spectra of Galactic cosmic rays (CRs) are not smooth power laws. PAMELA found that the proton and helium spectra from $1\,\rm{GV}$ to about $1\,\rm{TV}$ are harder at high rigidity than at low rigidity \cite{Adriani2011PAMELA}. AMS-02 confirmed this feature with much better statistics \cite{Aguilar2015Proton,Aguilar2015Helium}. At higher energies, CREAM-III, NUCLEON, DAMPE, and CALET reported additional structures in the multi-TeV range \cite{Yoon2017CREAM,Atkin2018NUCLEON,An2019DAMPEProton,Alemanno2021DAMPEHelium,Adriani2022CALETProton,Alemanno2024DAMPEPHe,DAMPE2026ChargeDependent}. These measurements raise a basic question: are the structures produced by large-scale Galactic propagation, or can a young nearby source contribute a visible component?

Several explanations have been proposed. The hardening may reflect a change in the rigidity dependence of Galactic diffusion, for example because of cosmic-ray-driven turbulence or spatially dependent transport \cite{Blasi2012Breaks,Tomassetti2012Hardening}. It may also be related to the injection or escape history of supernova remnants (SNRs) \cite{Ptuskin2013SNR}. A third possibility is that nearby discrete sources make a significant contribution over a finite energy interval \cite{Bernard2012Variance,Bernard2013Myriad,Thoudam2012NearbySNR,Liu2019NearbySource,Zhao2022GemingaSNR,Li2024CRAnisotropySpectra,Bhadra2025TeVBump,Evoli2021Stochastic}. This idea is attractive because it does not require the same feature to appear everywhere in the Galaxy. Its main difficulty is statistical. In a homogeneous diffusion model (HDM), Monte Carlo studies find that the possibility of a nearby source to significantly contribute the proton or helium spectrum is very low \cite{Evoli2021Stochastic}.

Such a conclusion depends on how particles transport near their sources. Escaping cosmic rays can amplify waves and reduce the local diffusion coefficient. Calculations of escape from SNRs show that particles may remain confined around the source for a finite time \cite{Gabici2009Clouds,Fujita2010Slow,Nava2016Nonlinear}. Gamma-ray observations also suggest that high-energy particles do not always diffuse as in the average interstellar medium. For example, LHAASO recently reported a giant ultra-high-energy gamma-ray bubble in the Cygnus X region, which can be described by continuous injection of high-energy protons into the ambient gas with suppressed diffusion \cite{LHAASOCygnusBubble2024}. Pulsar halos as observed by HAWC and LHAASO \cite{Abeysekara2017HAWC, Aharonian2021J0621} give another clue to slow diffusion around CR accelerators although other interpretations exist without suppressed diffusion \cite{Liu2019Anisotropic,Recchia2021TeVHalos,Yan2025MultiCoherence}.
These observations motivate models with locally inhibited transport around CR accelerators, i.e., the diffusion coefficient near the accelerator is smaller than the average in the interstellar medium (ISM). 

The physical effect of this two-zone diffusion model (2ZDM) is simple. In a HDM, particles spread rapidly after release and the flux at Earth is quickly diluted. In a 2ZDM, particles first cross an inner region with a smaller diffusion coefficient. Low-rigidity particles are delayed more strongly. Higher-rigidity particles can leave the inner region earlier and then diffuse through the outer region with a larger diffusion coefficient. As a result, the particles from a local source can reach Earth at a later time with extended duration, and temporal evolution of flux becomes more energy dependent. This effect is a redistribution of the arriving particles in time and energy.

To test this idea with modeling the contribution from the entire CR source population over the Galaxy as did in Ref.\cite{Evoli2021Stochastic}, one needs a fast propagation kernel. Direct numerical propagation is flexible, but it is inefficient for repeated Monte Carlo calculations if one must resolve both a small slow-diffusion region and a much larger outer region. We therefore derive an analytical series solution for a spherical two-zone diffusion problem. The inner region $0<r<r_1$ has diffusion coefficient $D_1$, the outer region $r_1<r<r_2$ has $D_2>D_1$, and both the density and diffusive flux are continuous at $r_1$. Once the eigenvalues are tabulated for given $\eta=D_2/D_1$ and $\beta=r_2/r_1$, the solution can be evaluated quickly for many source ages and distances.

With this kernel, we then follow the Monte Carlo source generation approach of Ref.~\cite{Evoli2021Stochastic} and calculate how often a random Galactic source population produces a dominant local source under the 2ZDM. We also compare instantaneous and continuous CR injection. In summary, we find that the 2ZDM can raise the probability of a local contribution comparable to the smooth background at 10\,TeV from the 0.1\% level (as found in the HDM \cite{Evoli2021Stochastic}) to the percent level. 
It shows that the possibility of an important contribution by a nearby source is significantly enhanced when near-source transport is assumed to be suppressed. Nevertheless, the $\sim 1\%$ level low probability further motivates us to use observed nearby pulsars as tracers of possible local SNRs, and investigate in which condition a local SNR can produce the bump-like feature around 10\,TeV seen in the data as suggested by some previous studies. 

The rest of the paper is organized as follows. Section~\ref{sec:analytical_solution} gives the two-zone Green's function used in the calculation. The derivation and numerical checks are given in the appendices. Section~\ref{sec:results} applies the solution to Monte Carlo source populations, and to cataloged nearby candidates. Section~\ref{sec:discussion} discusses the assumptions and limitations. Section~\ref{sec:conclusion} summarizes the main results and implications.

\section{Analytical two-zone solution}
\label{sec:analytical_solution}

This section gives the transport formula used below. The detailed derivation is presented in Appendix~\ref{app:two_zone_derivation}. We calculate the Green's function at a fixed energy, so $D_1$ and $D_2$ are the diffusion coefficients at that energy. We neglect energy losses, reacceleration, and nuclear spallation. This is adequate for the high-energy protons considered in this work.

The particle density is indicated by $C_1(r,t)$ in the inner region and $C_2(r,t)$ in the outer region. Here $r$ is the radial distance from the source and $t$ is the time since injection. They satisfy
\begin{align}
  \frac{\partial C_1}{\partial t}
  &=D_1\frac{1}{r^2}\frac{\partial}{\partial r}
  \left(r^2\frac{\partial C_1}{\partial r}\right),
  &&0<r<r_1,
  \label{eq:two_zone_inner_diffusion}
  \\
  \frac{\partial C_2}{\partial t}
  &=D_2\frac{1}{r^2}\frac{\partial}{\partial r}
  \left(r^2\frac{\partial C_2}{\partial r}\right),
  &&r_1<r<r_2 .
  \label{eq:two_zone_outer_diffusion}
\end{align}
We define
\begin{equation}
  \eta=\frac{D_2}{D_1},
  \qquad
  \beta=\frac{r_2}{r_1},
  \qquad
  m=\frac{\beta-1}{\sqrt{\eta}} .
  \label{eq:eta_beta_definition}
\end{equation}
The definitions of $D_1$, $D_2$, $r_1$ and $r_2$ were introduced in Section~\ref{sec:introduction}.
The source injects a total particle number $Q$ at the origin in a burst,
\begin{equation}
  C_1(r,0)=\frac{Q}{4\pi r^2}\delta(r).
  \label{eq:point_source_initial_condition}
\end{equation}
At the interface, both the density and the diffusive flux are continuous,
\begin{align}
  C_1(r_1,t)
  &=C_2(r_1,t),
  \\
  D_1\partial_r C_1(r_1,t)
  &=D_2\partial_r C_2(r_1,t).
  \label{eq:interface_conditions}
\end{align}
At the outer boundary we set
\begin{equation}
  C_2(r_2,t)=0 .
  \label{eq:outer_absorbing_boundary}
\end{equation}
This boundary closes the local Green's-function problem. It should not be confused with the boundary of the full Galactic halo.

The solution is a sum over eigenmodes. The eigenvalues $x_n>0$ are the roots of
\begin{equation}
\begin{aligned}
  \phi(x;\eta,\beta)
  ={}&\sqrt{\eta}\,x\sin x\cos(mx)
  \\
  &+\left[x\cos x+(\eta-1)\sin x\right]\sin(mx)=0 .
\end{aligned}
\label{eq:characteristic_equation}
\end{equation}
For each root,
\begin{equation}
  \Gamma_n=D_1\left(\frac{x_n}{r_1}\right)^2,
  \qquad
  {\cal D}_n=-\left.\frac{d\phi}{dx}\right|_{x=x_n} .
  \label{eq:decay_and_norm}
\end{equation}
For a burst source, the density is
\begin{align}
C_1(r,t)
&=\frac{Q}{2\pi r_1^2 r}
\sum_{n=1}^{\infty}
\frac{x_n^2\sin(mx_n)}
   {{\cal D}_n\sin x_n}
\sin\left(\frac{x_n r}{r_1}\right)
 e^{-\Gamma_n t},
\label{eq:time_solution_inner}
\\
C_2(r,t)
&=\frac{Q}{2\pi r_1^2 r}
\sum_{n=1}^{\infty}
\frac{x_n^2}{{\cal D}_n}
\sin\left[
\frac{x_n}{\sqrt{\eta}}
\left(\beta-\frac{r}{r_1}\right)
\right]
 e^{-\Gamma_n t} .
\label{eq:time_solution_outer}
\end{align}
Equation~\eqref{eq:time_solution_inner} applies for $0<r<r_1$, and Eq.~\eqref{eq:time_solution_outer} applies for $r_1<r<r_2$. For fixed $(\eta,\beta)$, the roots and coefficients only need to be computed once.

For a source injecting at a constant rate from $0$ to $t_0$, each factor $e^{-\Gamma_n t}$ is replaced by
\begin{equation}
  e^{-\Gamma_n t}\rightarrow
  {\cal T}_n(t;t_0)
  =\frac{1}{t_0}\int_0^{\min(t,t_0)}
  e^{-\Gamma_n(t-\tau)}\,d\tau .
  \label{eq:continuous_convolution}
\end{equation}
This gives
\begin{equation}
  {\cal T}_n(t;t_0)=
  \begin{cases}
  \dfrac{1-e^{-\Gamma_n t}}{\Gamma_n t_0}, & 0<t<t_0,\\[6pt]
  \dfrac{e^{-\Gamma_n(t-t_0)}-e^{-\Gamma_n t}}{\Gamma_n t_0}, & t>t_0 .
  \end{cases}
  \label{eq:continuous_time_factor}
\end{equation}

The above solution is for a fixed energy. We use an injection spectrum
\begin{equation}
Q(E)=q_0
\left(\frac{E}{10\,\rm{GeV}}\right)^{-\gamma_{\rm H}}
\exp\left(-\frac{E}{E_{\max}}\right).
\label{eq:injection_spectrum}
\end{equation}
Each source is assumed to inject a total cosmic-ray energy
$W_{\rm CR}=10^{50}\,\rm{erg}$.
The normalization constant $q_0$ is determined by requiring
\begin{equation}
W_{\rm CR}=\int_{1\,\rm{GeV}}^{\infty}
E\,Q(E)\,dE .
\label{eq:WCR_normalization}
\end{equation}

Following Evoli et al.~\cite{Evoli2021Stochastic}, we assume that diffusion is spatially homogeneous outside the local slow-diffusion region and depends only on particle energy. The diffusion coefficient is parameterized as
\begin{equation}
D_2(E) = D_0
\left(\frac{E}{\rm{GeV}}\right)^\delta
\left[
1+
\left(
\frac{E}{E_0}
\right)^{\Delta\delta/s}
\right]^{-s},
\label{eq:outer_diffusion_coefficient}
\end{equation}
where $D_0$ is the diffusion coefficient at $1\,\rm{GeV}$, $E_0$ is the break energy, $\Delta\delta$ determines the change of the diffusion slope, and $s$ controls the smoothness of the transition. This form accounts phenomenologically for the diffusion break inferred from the observed hardening of CR spectra around a few hundred $\rm{GV}$.

Therefore in the slow diffusion region:
\begin{equation}
D_1(E)=\frac{D_2(E)}{\eta}.
\label{eq:inner_diffusion_coefficient}
\end{equation}
For protons at the energies considered here, energy and rigidity can be used interchangeably up to an overall factor. The corresponding homogeneous diffusion solution is
\begin{equation}
C_{\rm HDM}(r,t, E) = \frac{Q(E)}{\left(4\pi D_2t\right)^{3/2}}
\exp\left[-\frac{r^2}{4 D_2t}\right].
\label{eq:homogeneous_solution}
\end{equation}
The differential intensity is related to the density by $\Phi(E)=\frac{c}{4\pi}C(E)$ for relativistic particles. This factor cancels in the flux ratios used below.

The parameters used in the calculations are summarized in Table~\ref{tab:Parameters}. Our solution can enhance the calculation accuracy and speed up the calculation significantly compared with another earlier analytical solution\cite{osipov2020energetic}. A detailed comparison is shown in Appendix~\ref{app:comparison_integral_solution}.

\begin{table}[!t]
\caption{Fiducial parameters adopted in this work.
The table summarizes the source injection parameters, diffusion parameters, and source-population parameters used throughout the calculations. The Galactic source-population and large-scale diffusion parameters are chosen to follow Ref.~\cite{Evoli2021Stochastic}, while $r_1$, $\eta$, $\beta$, and $t_0$ define the benchmark two-zone setup explored here.}
\label{tab:Parameters}
\begin{ruledtabular}
\begin{tabular}{lll}
Parameter & Description & Value \\
\hline
$W_{\rm CR}$ & Injected CR energy per source  & $10^{50}\,\rm{erg}$ \\
$\gamma_{\rm{H}}$ & Injection index & 2.30 \\
$E_{\max}$ & Cutoff energy & $500\,\rm{TeV}$ \\
$D_0$ & Diffusion coefficient & $2.2 \times 10^{28}\,\rm{cm^2\,s^{-1}}$ \\
$\delta$ & Diffusion index & 0.54 \\
$\Delta\delta$ & Diffusion-index change & 0.2 \\
$E_0$ & Break energy & $312\,\rm{GeV}$ \\
$s$ & Smoothing factor & 0.1 \\
$T_{\rm{sim}}$ & Simulation time & $10^8\,\rm{yr}$ \\
$\mathcal R_{\rm{SN}}$ & Supernova rate & $0.03\,\rm{yr^{-1}}$ \\
$h$ & Gas-disk half-height & $0.1\,\rm{kpc}$ \\
$R_\odot$ & Solar Galactocentric radius & $8.5\,\rm{kpc}$ \\
$(a,b)$ & Radial source parameters & $(1.9,5.0)$ \\
$r_1$ & Inner-zone radius & $0.05\,\rm{kpc}$ \\
$\eta$ & Diffusion coefficient ratio & 100 \\
$\beta$ & Radius ratio & 100 \\
$t_0$ & Injection time & $0.001\,\rm{Myr}$ \\
\end{tabular}
\end{ruledtabular}
\end{table}

\section{Results}
\label{sec:results}

\subsection{Effects of Two-Zone Diffusion}
\label{subsec:arrival_time}

In HDM, the time to reach distance $r$ is roughly controlled by $r^2/D_2(E)$. In 2ZDM, the propagation of CRs has two stages. Particles first cross the slow inner region, with a typical time $r_1^2/D_1(E)$, and then move through the faster outer region, with a typical time $(r-r_1)^2/D_2(E)$. The delay is larger for lower energies because both $D_1$ and $D_2$ increase with energy.

\begin{figure*}[!t]
\centering
\includegraphics[width=0.90\textwidth]{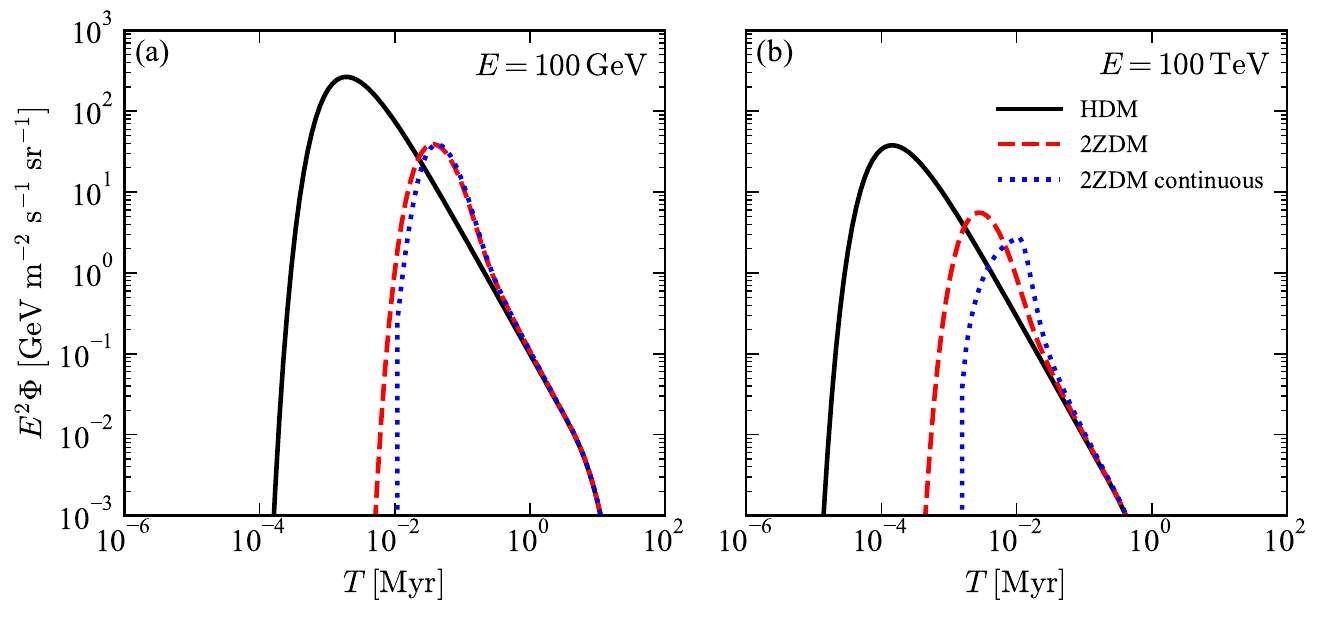}
\caption{Flux from a source at $r=0.1\,\rm{kpc}$ as a function of time calculated with parameters from Table~\ref{tab:Parameters} for different propagation models. Solid black, dashed red, and dotted blue curve represent the result under HDM with burst-like injection, 2ZDM with burst-like injection, and 2ZDM with continuous injection respectively. The left panel is for $100\,\rm{GeV}$, and the right panel is for $100\,\rm{TeV}$.}
\label{fig:Flux_Solution}
\end{figure*}

We quantify this delay effect by comparing three propagation scenarios: HDM with burst-like injection, 2ZDM with burst-like injection, and 2ZDM with finite-duration injection at a constant rate. Fig.~\ref{fig:Flux_Solution} shows the time evolution of the flux from a source at $r=0.1\,\rm{kpc}$. At $100\,\rm{GeV}$, the flux in HDM rises early and then declines. In both two-zone scenarios, the flux peaks later and stays above the homogeneous scenario for a longer time. At $100\,\rm{TeV}$, all curves move to earlier times, but the same delay is still visible. Finite-duration injection further broadens the peak. The slow-diffusion region therefore acts like a temporary reservoir of injected particles. It lowers the early peak, delays the release, and extends the duration of the flux peak. At later times, the flux enhancement in the 2ZDM weakens and the flux becomes almost identical to the prediction in the HDM. This is because particle injection has already stopped and even low-energy particles have largely escaped the inner slow-diffusion zone.

% \begin{figure*}[!t]
% \centering
% \includegraphics[width=0.98\textwidth]{figures/Phase_diagram.pdf}
% \caption{Ratio of the two-zone Green's function to the homogeneous Green's function for a source at $r=0.1\,\rm{kpc}$. The parameters are the same as in Fig.~\ref{Flux_Solution}. The slow inner region can increase the flux at some energies and times, while escape through the outer boundary suppresses the late high-energy flux.}
% \label{Phase_diagram}
% \end{figure*}

The importance of the two-zone effect depends on the diffusion contrast $\eta=D_2/D_1$, the size of the inner region, the source distance, and the source age. It is useful to introduce
\begin{equation}
\xi=\frac{r}{r_1},
\qquad
\tau=\frac{D_1 t}{r_1^2}.
\label{eq:dimensionless_variables}
\end{equation}
Here $\xi$ is the source-observer distance in units of the inner-zone radius, and $\tau$ is the source age in units of the diffusion time across the inner region. These variables allow different physical scales to be compared on the same diagram.

\begin{figure}[!t]
\centering
\includegraphics[width=1.0\linewidth]{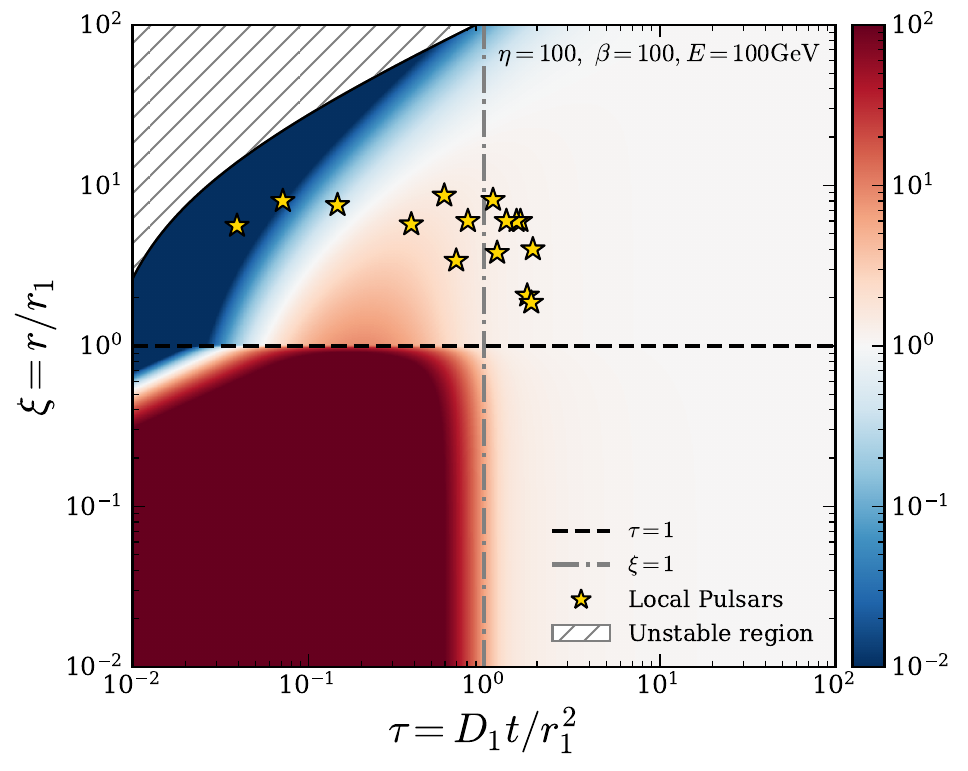}
\caption{Phase diagram of the flux ratio between the two-zone and homogeneous Green's functions in the plane of dimensionless distance $\xi=r/r_1$ and time $\tau=D_1t/r_1^2$. The hatched area marks the region where the physical flux at the observer is exponentially small and the finite-precision modal sum is not reliable. Yellow stars mark the selected nearby pulsars listed in Table~\ref{tab:Nearby_pulsars}.}
\label{fig:Phase_diagram}
\end{figure}

Fig.~\ref{fig:Phase_diagram} shows the flux ratio between the 2ZDM and the HDM in the $(\xi,\tau)$ plane. For an observer inside the inner region ($\xi<1$), the early flux is enhanced in the 2ZDM because particles are still confined near the source. For an observer outside the inner region ($\xi>1$), the early flux is suppressed because particles have not yet crossed the inner slow zone. The strongest observable deviation occurs near the transition region, roughly $\xi\gtrsim1$ and $\tau\sim1$, where particles are being released from the inner zone into the faster outer medium. At much larger $\tau$, the 2ZDM gradually approaches the HDM behavior, except for the effect of the finite outer boundary.

This diagram also motivates our selection criteria for local sources. For the benchmark $r_1=0.05\,\rm{kpc}$, sources within a few hundred pc correspond to $\xi$ of a few to ten. Ages below $1\,\rm{Myr}$ cover the range in which the delayed release can still affect multi-TeV particles. In the following subsections we therefore define local sources as objects with heliocentric distance $d<0.5\,\rm{kpc}$ and age $t<1\,\rm{Myr}$.

\subsection{Probability of an Important Contribution from a Local Source}
\label{subsec:local_probability}

We now investigate how often such a local source becomes the important contributor to the proton spectrum in a stochastic Galactic source population. We generate Monte Carlo realizations of SNRs using the Jelly model adopted by Evoli et al.~\cite{Evoli2021Stochastic}. More specifically, the radial distribution of SNRs is
\begin{equation}
P_R(R)\propto
\left(\frac{R}{R_\odot}\right)^a
\exp\left[
-b\frac{R-R_\odot}{R_\odot}
\right],
\end{equation}
where $R$ is the Galactocentric radius and
$R_\odot=8.5\,\rm{kpc}$ is the Galactocentric distance of the Sun. The vertical distribution is assumed to be Gaussian,
\begin{equation}
P_z(z)\propto
\exp\left(
-\frac{z^2}{2h^2}
\right).
\end{equation}

\begin{figure}[!t]
\centering
\includegraphics[width=0.9\linewidth]{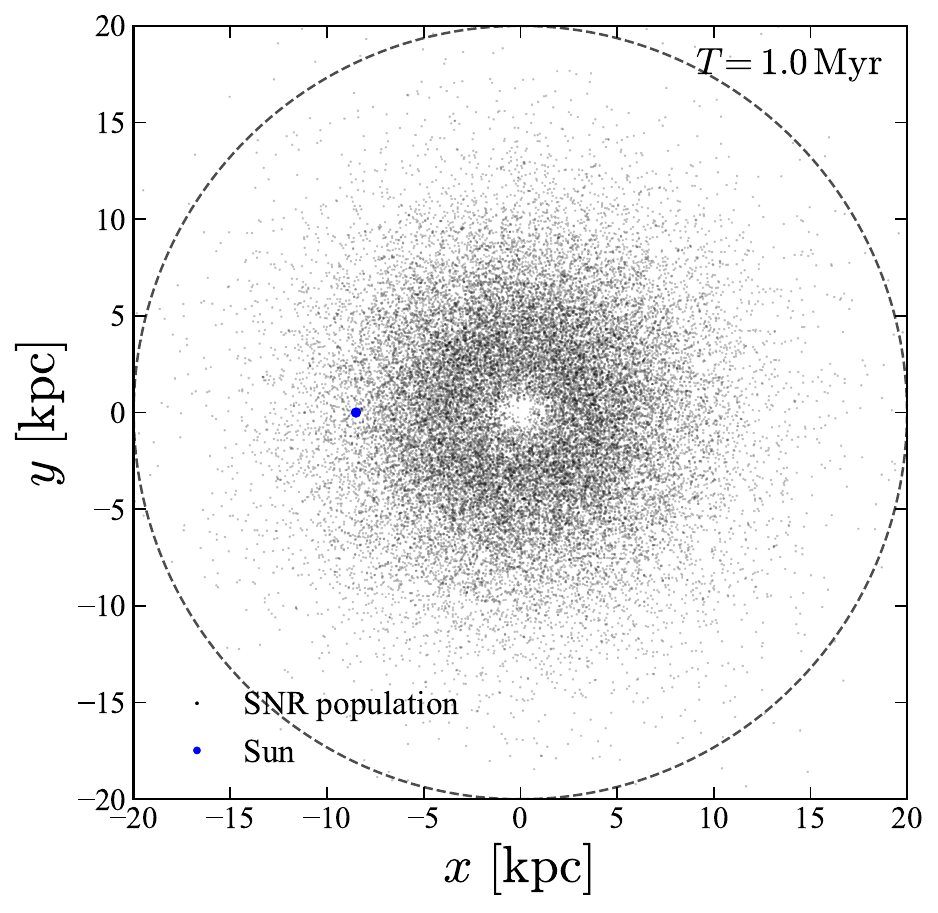}
\caption{One $1\,\rm{Myr}$ realization of the Jelly-model SNR distribution following Ref.~\cite{Evoli2021Stochastic}. The Sun is marked in blue at $R_\odot=8.5\,\rm{kpc}$.}
\label{fig:SNR_distribution}
\end{figure}

Source positions are generated by randomly sampling the distributions above, while source ages are assigned assuming a constant supernova rate of
$\mathcal R_{\rm{SN}} = 0.03\,\rm{yr^{-1}}$
during a total simulation time of
$T_{\rm{sim}}=10^8\,\rm{yr}$.
For the source population described above, there are about 10 sources on average that satisfy our local-source selection criteria. Fig.~\ref{fig:SNR_distribution} shows one realization of the resulting source population over a period of $1\,\rm{Myr}$.

\begin{figure*}[!t]
\centering
\includegraphics[width=1.0\textwidth]{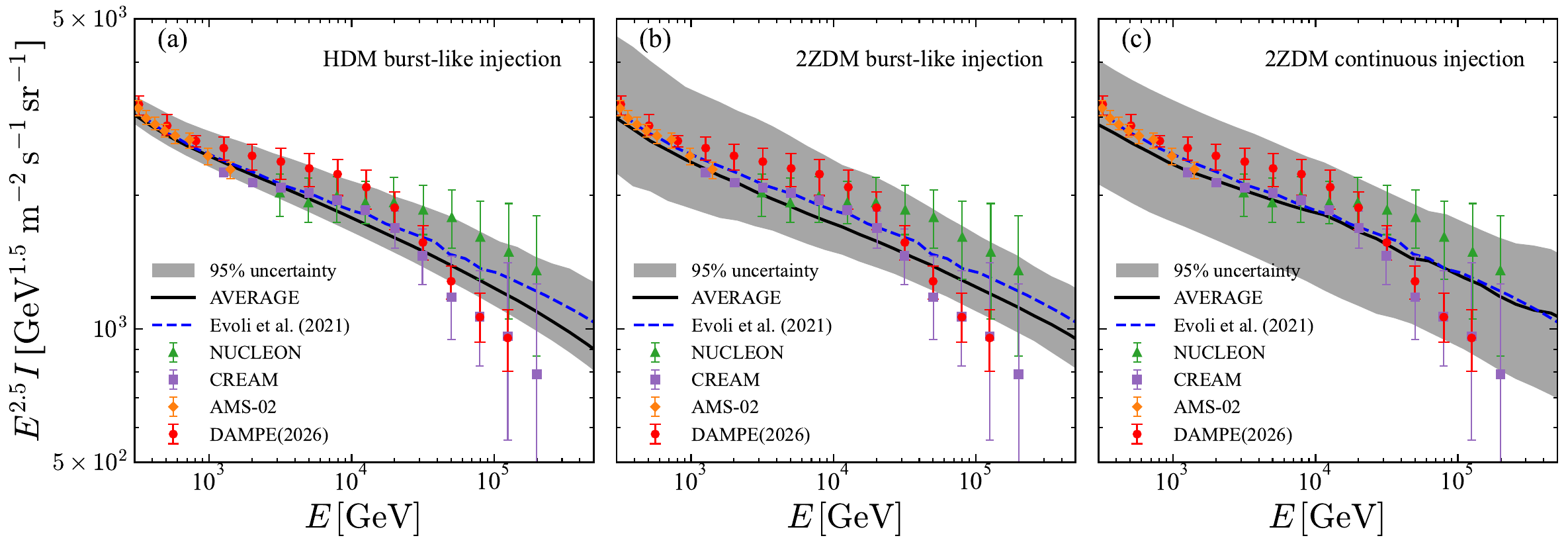}
\caption{Proton spectrum obtained from $10^4$ Monte Carlo realizations of the Galactic SNR population. The solid lines show the average flux, while the shaded regions represent the $95\%$ confidence interval, defined by the $2.5\%$ and $97.5\%$ percentiles of the flux distribution. The blue dashed line shows the result of Evoli et al.~\cite{Evoli2021Stochastic}. Panel (a) corresponds to the HDM. Panels (b) and (c) show the 2ZDM with $r_1=0.05\,\rm{kpc}$ and $\eta=\beta=100$, assuming burst-like injection and continuous injection ($t_0=0.001\,\rm{Myr}$), respectively. For all three figures the injection index is $\gamma_H=2.30$. The adopted CR energies per source are $W_{\rm CR}=10^{50}\,\rm{erg}$ per source in panel (a), $W_{\rm CR}=1.2\times10^{50}\,\rm{erg}$ in panel (b) and $W_{\rm CR}=1.8\times10^{50}\,\rm{erg}$ in panel (c). Data from DAMPE \cite{DAMPE2026ChargeDependent}, AMS-02 \cite{aguilar2015precision,aguilar2021alpha}, CREAM \cite{yoon2017proton}, and NUCLEON \cite{grebenyuk2019energy} are shown for comparison.}
\label{fig:Proton_spectrum}
\end{figure*}

In Fig.~\ref{fig:Proton_spectrum} we compare the predicted proton spectra with recent measurements. The source-population and propagation parameters are listed in Table~\ref{tab:Parameters}. For each propagation setup, $W_{\rm CR}$ is adjusted only to match the normalization of the observed average proton spectrum. This adjustment is needed because different propagation models give different residence times and therefore different absolute flux normalizations for the same injected energy. The required values remain of order $10^{50}\,\rm{erg}$, compatible with the standard SNR energy budget. This comparison shows that the adopted source population and propagation setups can reproduce the observed average spectrum. 

% The probability analysis below addresses a different question: how often does the strongest local source become comparable to the non-local background? Since this question is based on the ratio of two fluxes, a common rescaling of the injected energy of all sources cancels out. Therefore the different values of $W_{\rm CR}$ used for the illustrative spectra in Fig.~\ref{fig:Proton_spectrum} do not affect the distribution of the local-to-background ratio.

Then, to quantify the importance of nearby sources, we define
\begin{equation}
f \equiv
\frac{\Phi_{\rm loc}(10\,\rm{TeV})}
{\Phi_{\rm bkg}(10\,\rm{TeV})},
\end{equation}
where $\Phi_{\rm loc}$ is the flux from the strongest local source and $\Phi_{\rm bkg}$ is the total flux of all sources outside of the local-source selection criteria. Values of $f>1$ correspond to cases where a single nearby source contributes more flux than the background component. 

\begin{figure}[!t]
\centering
\includegraphics[width=1.0\linewidth]{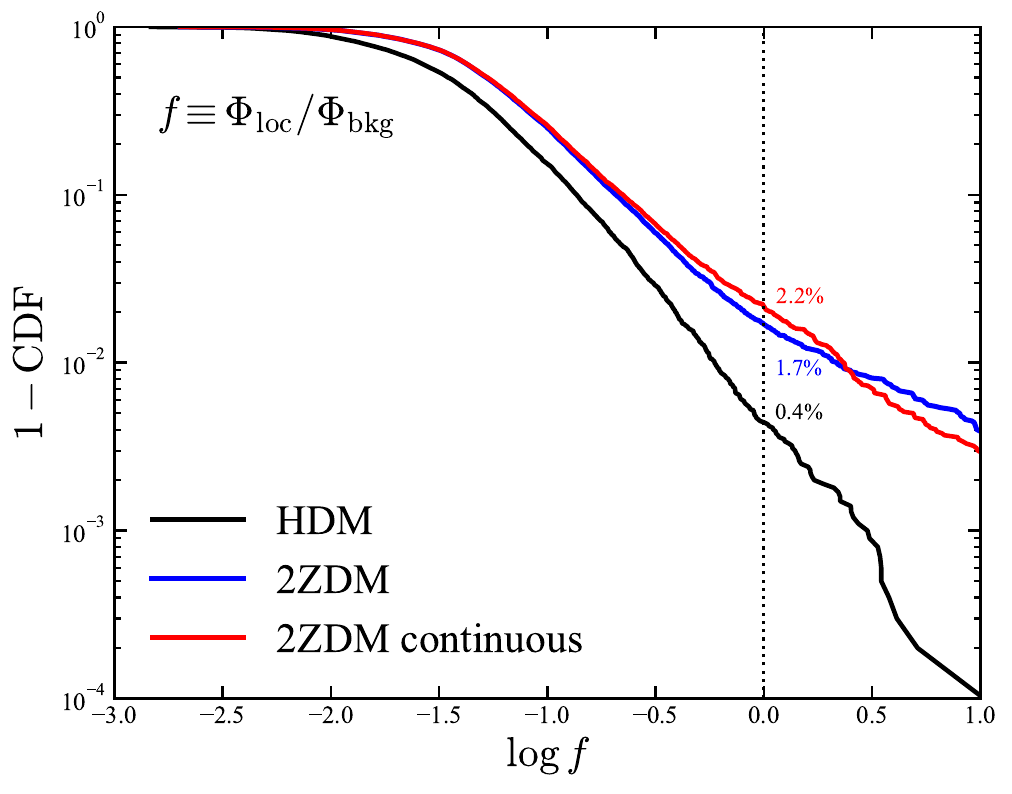}
\caption{The cumulative probability distribution of $f$ obtained from $10^5$ Monte Carlo realizations at $10\,\rm{TeV}$. For this probability calculation, the same injection parameters are used for all three propagation models, i.e., $W_{\rm CR}=10^{50}\,\rm{erg}$, $\gamma_{\rm H}=2.30$, and $E_{\max}=500\,\rm{TeV}$. At $f=1$, the probability is $0.4\%$ for HDM, $1.7\%$ for 2ZDM with burst-like injection, and $2.2\%$ for 2ZDM with continous injection.}
\label{fig:Probability_distribution}
\end{figure}

Fig.~\ref{fig:Probability_distribution} shows the cumulative probability distribution of $f$ obtained from $10^5$ Monte Carlo realizations at $10\,\rm{TeV}$. For this probability calculation, the same injection parameters are used for all three propagation models: $W_{\rm CR}=10^{50}\,\rm{erg}$, $\gamma_{\rm H}=2.30$, and $E_{\max}=500\,\rm{TeV}$. This choice isolates the effect of propagation on the local-to-background ratio. At $f=1$, the flux from the strongest local source becomes comparable to all the sources outside the local-source selection window. In the HDM, this occurs in only $0.4\%$ of realizations, consistent with the result of Ref.~\cite{Evoli2021Stochastic}. The probability increases to $1.7\%$ in the 2ZDM with burst-like injection and to $2.2\%$ when finite-duration injection is included.

Although the possibility of local-source dominance remains low, the slow-diffusion region increases its probability by a factor of about four to five. This result differs qualitatively from the conclusion reached under the HDM, where local-source explanations of TeV spectral features are generally considered unlikely. In the 2ZDM, such explanations remain uncommon but can no longer be regarded as statistically negligible.

\subsection{Contributions from Known Local Candidates}
\label{subsec:known_local_candidates}
Our result appears to be in tension with some previous works that have tried to explain the multi-TeV structure in the proton or all-particle spectrum by using specific nearby sources or local structures, such as Vela \cite{Thoudam2012NearbySNR,Bhadra2025TeVBump}, Geminga \cite{Liu2019NearbySource, Zhao2022GemingaSNR}, Monogem \cite{tang2022explanation}, Loop~I \cite{tang2022explanation} or other local sources. In these models, the local source parameters are adjusted to reproduce the observed feature. It is therefore useful to ask why/how these source-by-source studies can find important local contributions while the Monte Carlo calculation gives a small probability. 

There are several possible reasons for this difference. First, the real nearby source population may not be well represented by one random realization of the smooth Galactic SNR distribution. However, this possibility is difficult to test in a quantitative way, because we observe only one realization of the solar neighborhood. 
Second, different propagation models can change the CR arrival time and the flux normalization of the local source. A favored propagation model may enhance the predicted flux from certain source significantly. For example, Refs.\cite{Liu2019NearbySource, Zhao2022GemingaSNR} suggest a spatially dependent propagation (SDP) model with an effective diffusion coefficient of about $5\times 10^{28}\,\rm cm^2s^{-1}$ at 10\,TeV around several 100\,pc of the solar neighborhood, making the contribution of the Geminga SNR important. What's more, the injection spectrum parameters of the nearby source can affect the local contribution. Previous nearby-source models often employ a harder injection spectrum than that used in our Monte Carlo population study. To quantify the latter two effects, we keep the same two-zone transport framework, replace the random local sources by cataloged nearby SNR candidates, and study how the local contribution changes.

We use nearby pulsars as tracers of possible recent core-collapse SN events. The protons are assumed to be accelerated by the associated remnants even if their remnants are not observed.
Following the same source selection criteria, we select objects with heliocentric distance $d<0.5\,\rm{kpc}$ from Earth and characteristic age $t<1\,\rm{Myr}$ from the ATNF Pulsar Catalog \cite{Manchester2005ATNF} and additionally added Loop~I, as listed in Table~\ref{tab:Nearby_pulsars}. We show where these candidates lie in the $(\xi,\tau)$ plane in Fig.~\ref{fig:Phase_diagram}, and most of these candidates lie within or near the region where the two-zone effect is expected to be strongest.

\begin{table}[!t]
\caption{Basic properties of nearby pulsars used as representative local SNR candidates plus Loop~I. Distances are inferred from dispersion measurements with the YMW16 electron density model\citep{Yao2017}, and the age is based on characteristic ages of pulsars according to the ATNF Pulsar Catalog\cite{Manchester2005ATNF}. Distance and age of Loop~I is from Refs.\cite{galaxies6020056,Bhadra2025TeVBump}.}
\label{tab:Nearby_pulsars}
\centering
\begin{ruledtabular}
\begin{tabular}{cccc}
 Name & $d$ (kpc) & $\tau_c$ (Myr) \\
\hline
J0633+1746 (Geminga)  & 0.19 & 0.34 \\
B0656+14 (Monogem)  & 0.29 & 0.11 \\
B0833$-$45 (Vela)  & 0.28 & 0.01 \\
Loop~I & 0.17 & 0.20 \\
J0736$-$6304 & 0.10 & 0.51 \\
J0940$-$5428 & 0.38 & 0.04 \\
B0940$-$55  & 0.30 & 0.46 \\
B0941$-$56  & 0.41 & 0.32 \\
J0954$-$5430 & 0.43 & 0.17 \\
B0959$-$54  & 0.30 & 0.44 \\
B1055$-$52  & 0.09 & 0.54 \\
B1737$-$30  & 0.40 & 0.02 \\
J1741$-$2054 & 0.30 & 0.39 \\
B1742$-$30  & 0.20 & 0.55 \\
B1822$-$09  & 0.30 & 0.23 \\
\end{tabular}
\end{ruledtabular}
\end{table}

\begin{figure}[!t]
\centering
\includegraphics[width=1.0\linewidth]{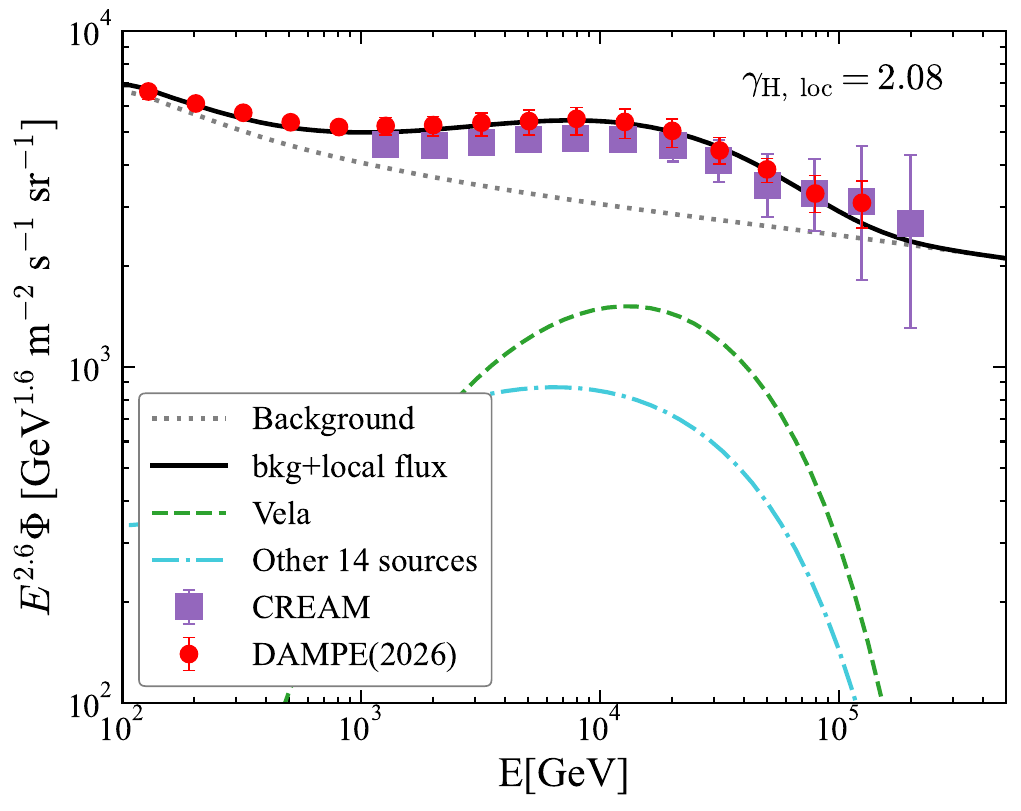}
\includegraphics[width=1.0\linewidth]{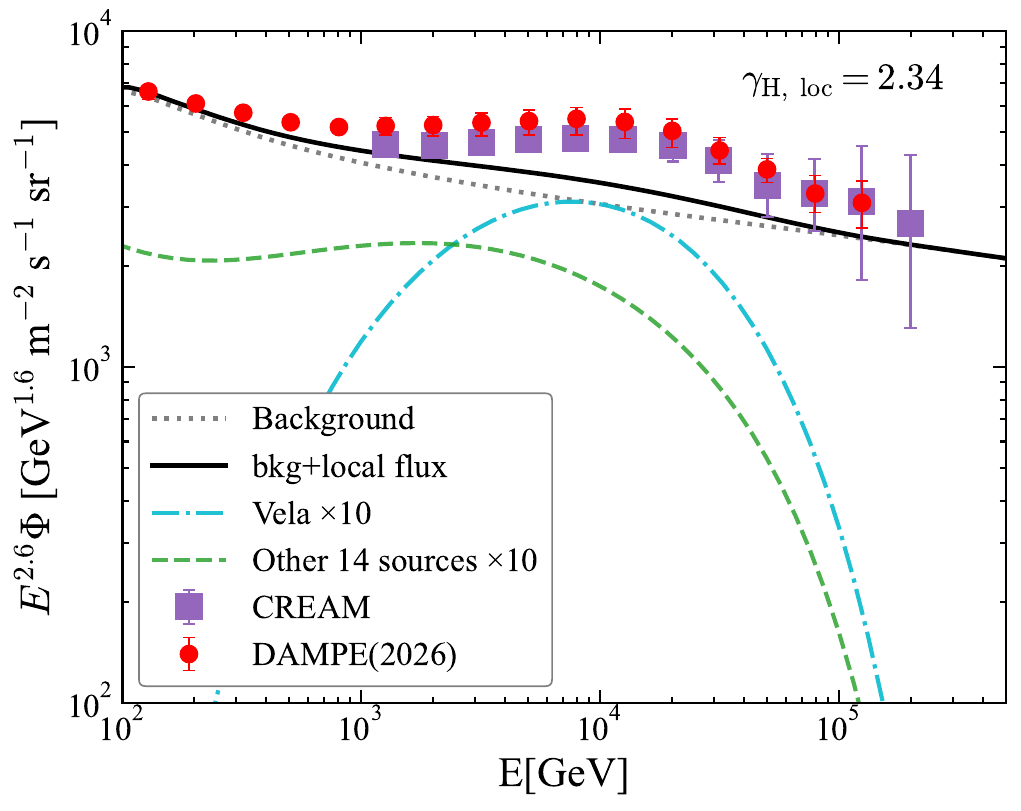}
\caption{
Proton spectrum obtained by adding the fifteen nearby candidates in Table~\ref{tab:Nearby_pulsars} to the stochastic Galactic background, with burst-like 2ZDM ($r_1=0.05\,\rm{kpc}$, $\eta=\beta=100$). The background is defined as the average flux over $10^4$ realizations after excluding sources with heliocentric distance $d<0.5\,$kpc. All sources are assumed to inject $W_{\rm CR}=10^{50}\,\mathrm{erg}$.
Upper panel: $\gamma_{\rm H,bkg}=2.34$, $E_{\max,bkg}=500\,\mathrm{TeV}$ for the background population, and $\gamma_{\rm H,loc}=2.08$, $E_{\max,loc}=50\,\mathrm{TeV}$ for the nearby candidates. Lower panel: same as the upper panel, except $\gamma_{\rm H,loc}=2.34$. Nearby-source fluxes in the lower panel are multiplied by 10 for visibility. Data are from DAMPE \cite{DAMPE2026ChargeDependent} and CREAM \cite{yoon2017proton}.
}
\label{fig:DAMPE_fit}
\end{figure}

Fig.~\ref{fig:DAMPE_fit} shows the proton spectrum obtained by adding the selected nearby candidates to a stochastic Galactic background. The background component is constructed using the same source-population model as in Fig.~\ref{fig:Proton_spectrum}. We generate $10^4$ realizations of the Galactic SNR population, exclude sources with heliocentric distance $d<0.5\,$kpc, and define the background flux as the average two-zone propagated contribution from the remaining sources.
For the background sources we adopt $W_{\rm CR}=10^{50}\,\rm{erg}$, $\gamma_{\rm H,bkg}=2.34$, and $E_{\rm max,bkg}=500\,\rm{TeV}$. The local component is computed as the sum of the nearby candidates listed in Table~\ref{tab:Nearby_pulsars}, also propagated with 2ZDM. For these nearby sources we adopt $W_{\rm CR}=10^{50}\,\rm{erg}$, $\gamma_{\rm H,loc}=2.08$, and $E_{\rm max,loc}=50\,\rm{TeV}$.

The upper panel of Fig.~\ref{fig:DAMPE_fit} shows the resulting spectrum. Owing to the harder injection spectrum of the nearby sources, their contribution remains significant at multi-TeV energies. Among all candidates, Vela provides the largest individual contribution and dominates the local component above several TeV. The total spectrum develops a feature around $10\,\rm{TeV}$ and shows improved agreement with the latest DAMPE proton data \cite{DAMPE2026ChargeDependent}. For comparison, the lower panel shows the result obtained when the nearby sources are assigned the same injection index as the background population($\gamma_{\rm H,loc}=\gamma_{\rm H,bkg}=2.34$), while all other parameters are kept unchanged ($W_{\rm CR}=10^{50}\,\rm{erg}$ and $E_{\rm max,loc}=50\,\rm{TeV}$). Since a larger fraction of the injected energy is carried by lower-energy particles, the contribution at multi-TeV energies is strongly suppressed and becomes negligible compared with the background. For visibility, the nearby-source fluxes shown in the lower panel have been multiplied by a factor of 10.

\begin{figure*}[!t]
\centering
\includegraphics[width=1.0\textwidth]{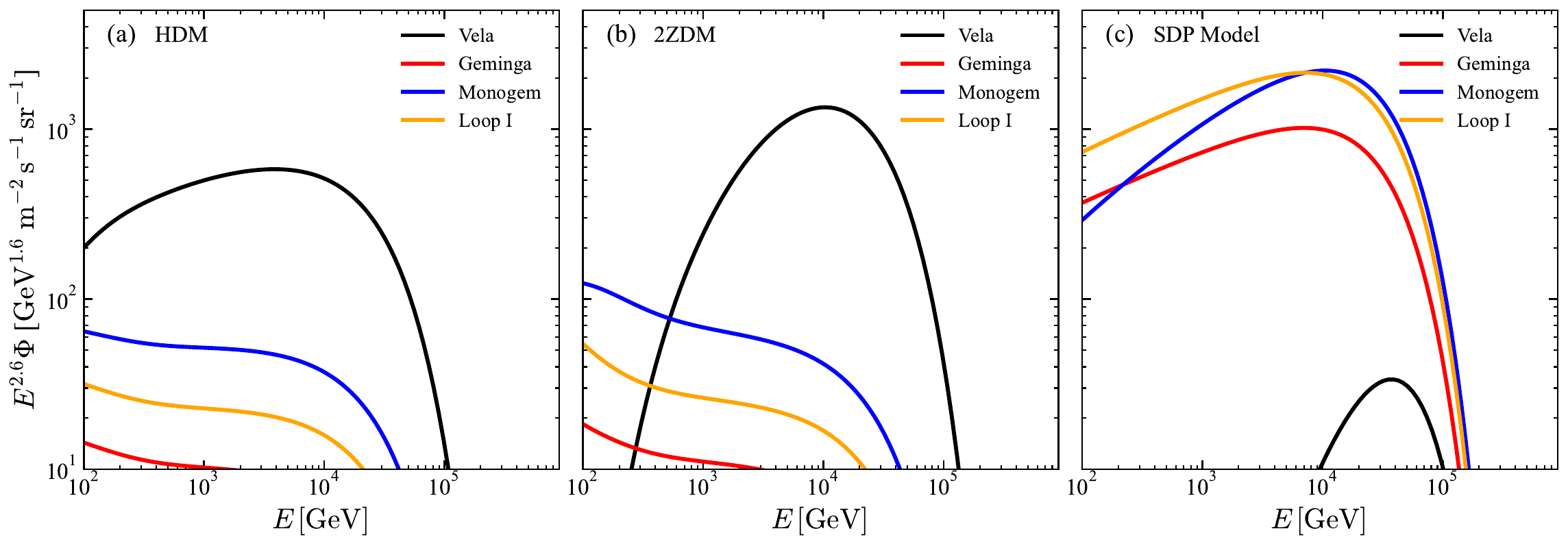}
\caption{
Proton fluxes from Vela, Geminga, Monogem, and Loop~I for different propagation models. 
The panels show (a) HDM, (b) 2ZDM, and (c) SDP adopted by Liu et al.~\cite{Liu2019NearbySource} and Zhao et al.~\cite{Zhao2022GemingaSNR}. 
All candidates are assigned the same injection spectrum $W_{\rm CR}=10^{50}\,\rm{erg}$, $\gamma_{\rm H}=2.08$, and $E_{\max}=50\,\rm{TeV}$ with burst-like injection.
}
\label{fig:Compare_propagation}
\end{figure*}

The propagation model for local sources also has an important influence on the predicted local contribution, as shown in Fig.~\ref{fig:Compare_propagation}. In this figure, we compare the fluxes predicted for Vela, Geminga, Monogem, and Loop~I under HDM, 2ZDM, and SDP. For the SDP case, we adopt the parameter set used by Zhao et al.~\cite{Zhao2022GemingaSNR}. The injection spectra are identical in all cases, so the differences arise solely from CR transport. 
%The comparison demonstrates that the identification of the dominant nearby source depends largely on the transport model, especially the effective diffusion coefficient. At the Solar position, the diffusion coefficient in the HDM and in the outer region of the 2ZDM is $D=2.62\times10^{29}\,\rm{cm^2\,s^{-1}}$ at $100\,\rm{GeV}$ and $1.59\times10^{30}\,\rm{cm^2\,s^{-1}}$ at $10\,\rm{TeV}$. By contrast, the SDP model gives a much smaller local diffusion coefficient, $D=2.48\times10^{28}\,\rm{cm^2\,s^{-1}}$ at $100\,\rm{GeV}$ and $5.35\times10^{28}\,\rm{cm^2\,s^{-1}}$ at $10\,\rm{TeV}$. 
A source can make an important contribution when $d^2/(D_{\rm eff}t_{\rm age})\sim 1$. In the HDM, multi-TeV particles spread quickly after leaving the source, so the flux at Earth is small unless the source is very close, very young, or very powerful. In the 2ZDM, the inner region keeps multi-TeV particles near the source for some time and releases them later, which can enhance the local contribution. In the SDP model, slow diffusion extends over a larger volume around the solar neighborhood, so particles take longer time to arrive at Earth after release. This allows older and more distant sources such as the Gemigna SNR to make important contributions. This is consistent with Ref.\cite{Zhao2022GemingaSNR}. However, interestingly,we found that the Monogem SNR and Loop~I make higher contribution than the Geminga SNR  in the SDP case, which was not addressed by Ref.~\cite{Zhao2022GemingaSNR}.

\section{Discussion}
\label{sec:discussion}

We note that the parameters in Table~\ref{tab:Parameters} are not unique. They are benchmark values based on previous studies, and this does not mean that every source should have the same injection index and cutoff energy. The same is true for $r_1$ and $\eta$. We have tried varying the injection index for each SNR and found that it can fit the observed proton energy spectrum even when it fluctuates by 0.2 around 2.3. However, when repeating the Monte Carlo probability calculation in Section~\ref{subsec:local_probability} with the injection index fluctuating, the result did not show significant changes, and our conclusion remains unchanged. Besides, although the proton spectrum cutoff is taken to be 7\,PeV in the SNRs generated in our Monte Carlo calculation, this does not mean that we suggest SNRs to be solely responsible for measured CRs from GeV to multi-PeV. Recent LHAASO observations suggest that microquasars\cite{LHAASO_microquasar, LHAASO_X3}, young massive stellar clusters\cite{LHAASOCygnusBubble2024}, and young PWNe may be PeV particle accelerators\cite{LHAASO_Crab, LHAASO_J1849}, so CRs around and beyond the ``knee'' may be contributed by one or more classes of these sources instead of SNRs. 

The use of pulsars as source tracers is another approximation. First, the characteristic ages and DM-based distances as provided in Table~\ref{tab:Nearby_pulsars} may have large uncertainties. Second, pulsars only trace core-collapse SN events but not thermonuclear SN events. Besides, we may also miss some core-collapse SNRs if the radiation beam of their associated pulsars do not point to us. Nevertheless, this would not change our conclusion qualitatively since we aim to investigate the influence of the source age and distance, and propagation model. However, if we aim to pinpoint local contributors of the bump-like spectral structure in the future study,  we need to consider the nearby contribution based on a more complete nearby samples, probably by combining pulsar catalogs, SNR catalogs, local gas maps, and gamma-ray data.

It is also worth noting that, a strong local CR source should affect not only the proton spectrum, but also the helium spectrum, composition-dependent spectral features, secondary-to-primary ratios, diffuse gamma-ray emission from nearby gas, and the CR anisotropy. The anisotropy constraint is especially important since it depends on the spatial gradient of the local CR density. However, anisotropy predictions are also sensitive to the same transport assumptions discussed above, especially anisotropic diffusion and the structure of the local magnetic field. Therefore a robust test of the local source interpretation requires joint modeling of spectra, composition, anisotropy, and gamma-ray data within a consistent description of the local ISM. On the other hand, the nearby-source contribution would be more prominent at higher energies, because the number of sources capable of producing higher-energy CRs is supposed to be fewer. Future observations on fine CR spectral structure in $\sim 1-100$\,PeV range by large air-shower arrays such as LHAASO, GRAND\cite{GRAND_WB}, Pierre Auger Observatory\cite{Auger2025}, Telescope Array\cite{TA2018} and etc may also provide crucial test on the local-source effect (although the ``local'' volume defined for these high-energy particles should be much larger than that for TeV particles discussed above). 

% In Fig.~\ref{fig:DAMPE_fit}, the combined contribution of the fourteen nearby candidates becomes comparable to the background flux, with Vela providing the dominant contribution. This does not contradict the low probabilities obtained in Section~\ref{subsec:local_probability}, since the two calculations adopt different injection spectra. In the Monte Carlo study, all sources were assigned $\gamma_{\rm H}=2.30$, whereas the nearby candidates in Fig.~\ref{fig:DAMPE_fit} were assumed to have a harder spectrum, $\gamma_{\rm{H,loc}}=2.08$.

% Hard injection spectra are commonly invoked in nearby-source interpretations of the TeV proton spectrum. They are also consistent with the expectation that young supernova remnants accelerate particles with spectra close to $E^{-2}$, while older remnants produce progressively softer spectra. A harder injection spectrum therefore increases the contribution of nearby sources at multi-TeV energies.

\section{Conclusion}
\label{sec:conclusion}

In summary, we studied the contribution of local sources to the Galactic CR spectrum and its dependence on transport assumptions. For this purpose we derived an analytical Green's function for an isotropic two-zone diffusion model, which allow us to do fast calculation of CR distribution. In this model, the inner slow region delays particle escape and changes the time and energy dependence of the flux observed at Earth. 

Using Monte Carlo realizations of Galactic SNR populations, we found that inhibited near-source transport increases the chance of an important local contribution. At $10\,\rm{TeV}$, the probability that the strongest local source contribution becomes comparable to that of background sources rises from $0.4\%$ in the homogeneous diffusion model to $1.7$--$2.2\%$ in the two-zone diffusion models. This increase is significant, but the absolute probability remains small. Therefore a dominant local source is not expected in most realizations if local sources share the same injection properties as the average Galactic population. 

We also studied cataloged nearby candidates, using nearby pulsars as tracers of possible recent SNRs. This calculation shows that known local candidates can produce a feature around $10\,\rm{TeV}$ only under additional specific assumptions. In our benchmark example, a hard local injection spectrum is required, which is consistent with the findings in previous studies. If the local injection index is close to that of the background population, the local contribution becomes negligible. The predicted flux also changes strongly when the same candidates are propagated with homogeneous diffusion, localized two-zone diffusion, or spatially dependent propagation. Thus the importance of a given nearby source is controlled not only by its age and distance, but also by the assumed magnetic environment between the source and the solar system which would determine the transport properties of injected particles. 

Our main conclusion is therefore that local source interpretations of multi-TeV CR spectral structures are highly model dependent. A nearby source can explain a spectral feature for favorable source parameters and transport conditions, but such a fit is not unique and should not be interpreted as direct evidence for that source. To establish a robust local source origin, one must constrain the particle injection of the candidate source, and the particle propagation properties from ambient medium of the accelerator to the solar neighborhood. Future studies combining CR spectra and composition with anisotropy measurements, gamma-ray observations of nearby gas and accelerators in the modeling would provide important insights into this issue.

\begin{acknowledgments}
This work is funded by National SKA Program of China under grant No.~2025SKA0110104, National Natural Science Foundation of China under grant No.~12393852, and Basic Research Program of Jiangsu under grant No.~BK20250059. 
\end{acknowledgments}

\appendix

\section{Derivation of the two-zone Green's function}
\label{app:two_zone_derivation}

\begin{figure}[!t]
\includegraphics[width=1.0\linewidth]{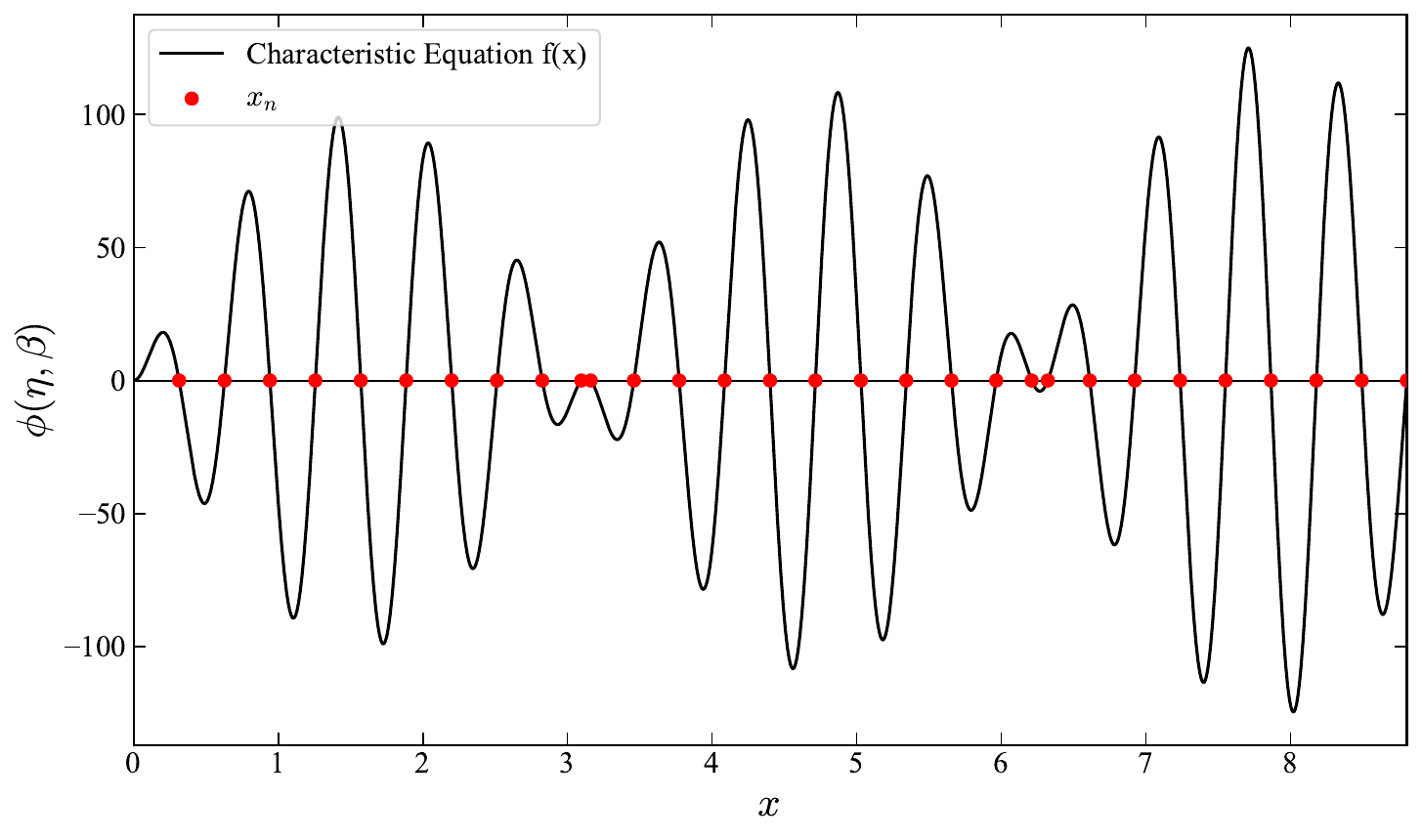}
\caption{Numerical roots of the characteristic equation for $\eta=100$ and $\beta=100$.}
\label{fig:Numerical_roots}
\end{figure}

Here we derive the series solution used in Sec.~\ref{sec:analytical_solution}. We define the Laplace transform
\begin{equation}
  \widehat C_k(r,s)=\int_0^\infty e^{-st}C_k(r,t)\,dt,
  \qquad k=1,2,
\end{equation}
and use
\begin{equation}
  \lambda=r_1\sqrt{\frac{s}{D_1}},
  \qquad
  \rho=\frac{r}{r_1},
  \qquad
  m=\frac{\beta-1}{\sqrt{\eta}} .
\end{equation}
For $r>0$ the transformed equation is homogeneous. The point source fixes the singular part of the inner solution,
\begin{equation}
  \widehat C_1(r,s)\simeq \frac{Q}{4\pi D_1r},
  \qquad r\rightarrow0^+ .
\end{equation}
We write the solution as
\begin{align}
  \widehat C_1(r,s)
  &=\frac{1}{r}\left[
    \frac{Q}{4\pi D_1}e^{-\lambda\rho}
    +A(s)\sinh(\lambda\rho)
  \right],
  \\
  \widehat C_2(r,s)
  &=\frac{B(s)}{r}
  \sinh\left[\frac{\lambda}{\sqrt{\eta}}(\beta-\rho)\right].
\end{align}
The second line satisfies the absorbing boundary at $r_2$. Applying the two interface conditions gives
\begin{align}
  B(s)&=\frac{Q}{4\pi D_1}
  \frac{\lambda}{\Phi(\lambda;\eta,\beta)},
  \\
  A(s)&=\frac{Q}{4\pi D_1}e^{-\lambda}
  \frac{(\lambda+1-\eta)\sinh(m\lambda)
  -\sqrt{\eta}\lambda\cosh(m\lambda)}
  {\Phi(\lambda;\eta,\beta)},
\end{align}
where
\begin{equation}
\begin{aligned}
  \Phi(\lambda;\eta,\beta)
  ={}&\left[\lambda\cosh\lambda+(\eta-1)\sinh\lambda\right]
  \sinh(m\lambda)
  \\
  &+\sqrt{\eta}\lambda\sinh\lambda\cosh(m\lambda).
\end{aligned}
\end{equation}
The poles are obtained by setting $\lambda=ix$, which gives the characteristic equation in Eq.~\eqref{eq:characteristic_equation}. Fig.~\ref{fig:Numerical_roots} plots the numerical roots of the characteristic equation for $\eta=100$ and $\beta=100$.

A useful check is the homogeneous infinite limit. For $\eta=1$, the root equation becomes
\begin{equation}
  \phi(x;1,\beta)=x\sin(\beta x),
\end{equation}
so $x_n=n\pi/\beta$ and $k_n=x_n/r_1=n\pi/r_2$. Taking $r_2\rightarrow\infty$ turns the sum into
\begin{equation}
  C(r,t)=\frac{Q}{2\pi^2 r}
  \int_0^\infty k\sin(kr)e^{-Dk^2t}\,dk .
\end{equation}
The integral gives
\begin{equation}
  C(r,t)=\frac{Q}{(4\pi Dt)^{3/2}}
  \exp\left(-\frac{r^2}{4Dt}\right),
\end{equation}
which is the standard three-dimensional Green's function. This check fixes the normalization of the point source and the root condition.

\section{Minimum number of eigenmodes for convergence}

The analytical solution is expressed as an infinite sum over eigenmodes and must therefore be truncated in practical calculations. The required number of roots depends on both the desired accuracy and the minimum source age considered in the calculation.

The temporal factor of the $n$th eigenmode is
\begin{equation}
\exp\left[
-D_1
\left(
\frac{x_n}{r_1}
\right)^2
t
\right].
\end{equation}

To achieve an accuracy corresponding to $k\sigma$, the contribution from neglected higher-order modes should be smaller than the target threshold
\begin{equation}
\exp\left(
-\frac{k^2}{2}
\right).
\end{equation}

For the youngest source age $t_{\min}$ included in the calculation, this requires
\begin{equation}
\exp\left[
-D_1
\left(
\frac{x_{\max}}{r_1}
\right)^2
t_{\min}
\right]
\le
\exp\left(
-\frac{k^2}{2}
\right),
\end{equation}

which gives
\begin{equation}
x_{\max}
\ge
\frac{kr_1}
{\sqrt{2D_1t_{\min}}}.
\end{equation}

For large mode number, the roots become approximately equally spaced,
\begin{equation}
\Delta x
\approx
\frac{\pi\sqrt{\eta}}
{\beta}.
\end{equation}

The minimum number of roots required is therefore
\begin{equation}
N_{min} \approx \frac{x_{max}}{\Delta x} = \frac{k \cdot r_1 \cdot \beta}{\pi \sqrt{2 D_1 t_{min} \eta}}.
\end{equation}

Table~\ref{tab:Truncation_parameters} lists the estimated values of $x_{\max}$ and $N_{\min}$ for several representative choices of $t_{\min}$ and accuracy levels. For the parameter range used in this work, a few thousand roots are generally sufficient to achieve convergence.

\begin{table}[htbp]
\centering
\caption{Required truncation parameters for different starting times and accuracy levels.}
\label{tab:Truncation_parameters}
\begin{tabular}{ccccccc}
\hline
 & \multicolumn{1}{c}{3$\sigma$ accuracy} 
 & \multicolumn{1}{c}{5$\sigma$ accuracy} 
 & \multicolumn{1}{c}{7$\sigma$ accuracy} \\
$t_{\min}$ & $N$ & $N$ & $N$ \\
\hline
$10^{-4}$ Myr
& $  5000$
& $  10000$
& $  15000$ \\
$10^{-3}$ Myr
& $  2000$
& $  4000$
& $  5000$ \\
$10^{-2}$ Myr
& $  500$
& $  1000$
& $  1500$ \\
$10^{-1}$ Myr
& $  200$
& $  400$
& $  500$ \\
\hline
\end{tabular}
\end{table}

\section{Comparison with the previous integral solution}
\label{app:comparison_integral_solution}

In the main text we use the eigenmode series solution because it is both accurate and efficient for repeated evaluations. Here we make a quantitative comparison with the analytical two-zone integral propagator of Osipov et al.~\cite{osipov2020energetic}. To isolate the spatial propagation kernel, we neglect energy losses in both calculations and compare them at fixed energy. The integral solution assumes an effectively infinite outer region, whereas our series solution uses a finite absorbing boundary at $r_2=\beta r_1$. We therefore compare the two solutions in the time range where the finite outer boundary does not affect the result.

In the notation used here, the inner diffusion coefficient is $D_1$, the outer diffusion coefficient is $D_2$, and the transition radius is $r_1$. For fixed energy and diffusion time $t$, the integral solution can be written as\cite{osipov2020energetic}
\begin{equation}
C_{\rm int}(r,t)=Q\,\mathcal H(r,t),
\end{equation}
where
\begin{equation}
\begin{split}
&\mathcal{H}(r,t)=\int_0^\infty d\psi\,
\frac{\xi e^{-\psi}}{\pi^2\lambda_1^2\left[A^2(\psi)+B^2(\psi)\right]}\\
&\times \left\{
\begin{array}{ll}
\dfrac{1}{r}\sin\left(2\sqrt{\psi}\dfrac{r}{\lambda_1}\right), & 0<r<r_1,\\[8pt]
\dfrac{A(\psi)}{r}\sin\left(2\sqrt{\psi}\dfrac{\xi r}{\lambda_1}\right)
+\dfrac{B(\psi)}{r}\cos\left(2\sqrt{\psi}\dfrac{\xi r}{\lambda_1}\right), & r\ge r_1.
\end{array}
\right.
\end{split}
\label{eq:osipov_integral_kernel}
\end{equation}
Here
\begin{equation}
\lambda_1=\sqrt{4D_1t},
\qquad
\xi=\sqrt{\frac{D_1}{D_2}}=\frac{1}{\sqrt{\eta}},
\qquad
\chi=2\sqrt{\psi}\frac{r_1}{\lambda_1},
\end{equation}
and
\begin{equation}
\begin{aligned}
A(\psi)={}&\xi\cos\chi\cos(\xi\chi)
+\sin\chi\sin(\xi\chi)
\nonumber\\
&+\frac{1}{\chi}\frac{1-\xi^2}{\xi}\sin\chi\cos(\xi\chi),
\end{aligned}
\end{equation}
\begin{equation}
B(\psi)=\frac{\sin\chi-A(\psi)\sin(\xi\chi)}{\cos(\xi\chi)} .
\end{equation}
We evaluate Eq.~\eqref{eq:osipov_integral_kernel} numerically using Gauss--Laguerre quadrature, which is appropriate for the weight factor $e^{-\psi}$.

\begin{figure}[htbp]
\centering
\includegraphics[width=1.0\linewidth]{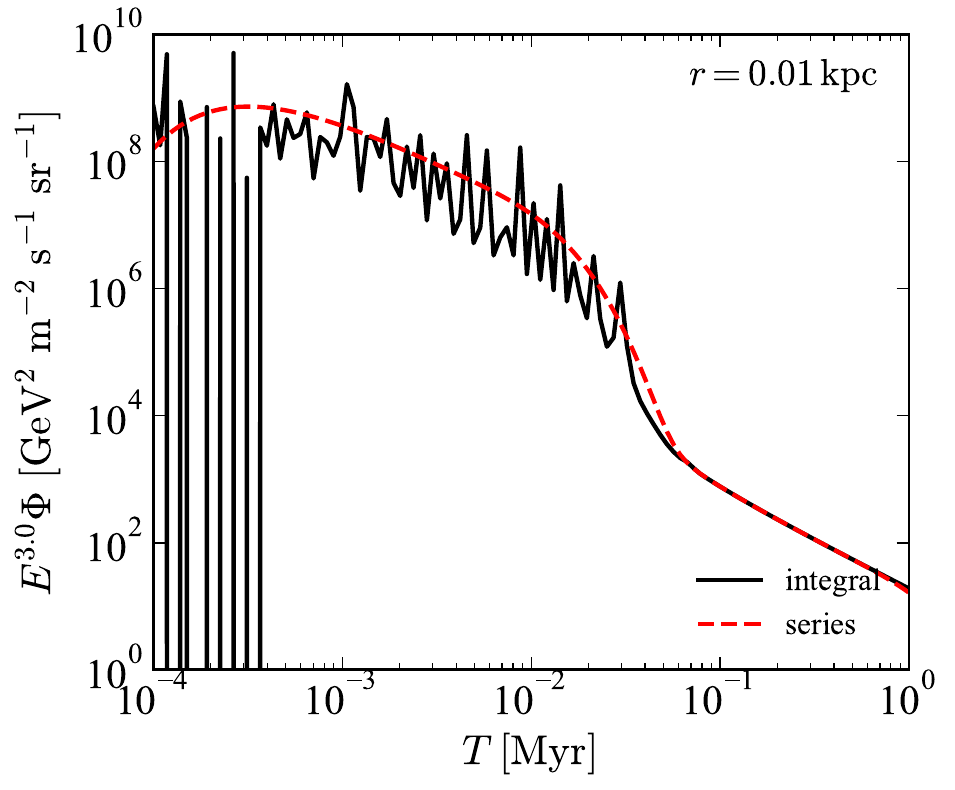}
\includegraphics[width=1.0\linewidth]{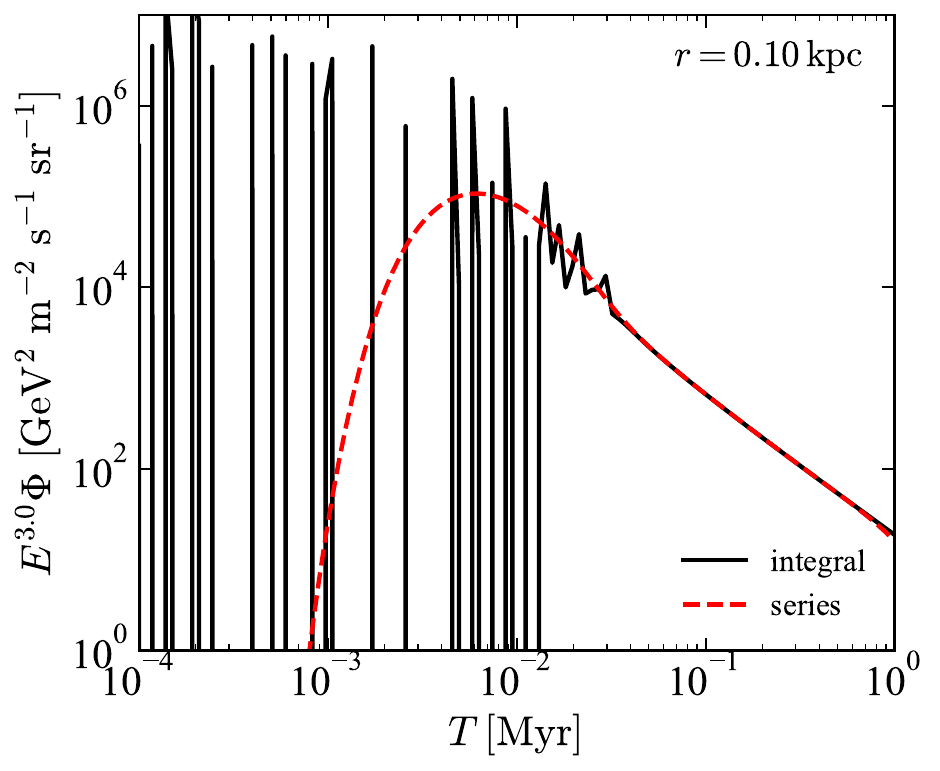}
\caption{Comparison between the integral solution by Osipov et al.~\cite{osipov2020energetic} (solid black curves) and the eigenmode series solution derived in this work (dashed red curves). Energy losses are neglected in both calculations. The curves show $E^{3}\Phi$ at $E=10\,\rm{TeV}$ for an observer inside the slow-diffusion region ($r=0.01\,\rm{kpc}$) in the upper panel and outside it ($r=0.1\,\rm{kpc}$) in the lower panel. The benchmark parameters are $r_1=0.05\,\rm{kpc}$ and $\eta=\beta=100$. The series solution uses 500 roots. The integral solution uses 160 Gauss-Laguerre integration nodes on the integral variable $\psi$.}
\label{fig:compare_solution_stability}
\end{figure}

For the benchmark parameters $r_1=0.05\,\rm{kpc}$ and $\eta=\beta=100$, Appendix~B shows that 500 roots are already sufficient for source ages of order $10^{-2}\,\rm{Myr}$ at the $3\sigma$ level. We therefore use 500 roots in the comparison below. Fig.~\ref{fig:compare_solution_stability} compares the time evolution of $E^3\Phi(E)$ at $E=10\,\rm{TeV}$ for two observer distances, $r=0.01\,\rm{kpc}$ inside the slow-diffusion region and $r=0.1\,\rm{kpc}$ outside it. The integral solution is evaluated with 160 Gauss--Laguerre quadrature points. The eigenmode series gives a smooth and stable result with 500 roots and agrees with the integral representation in the regime where the two formulations describe the same physical problem.

Our solution can also save computational time, which can be illustrated with a representative calculation of a population of sources under the 2ZDM. We evaluate the propagator on a grid of source ages and distances, covering $(10^{-4}-1)\,\mathrm{Myr}$ with 50 logarithmically spaced bins and $(10^{-2}-1)\,\mathrm{kpc}$ with 20 logarithmically spaced bins. For a population of 100 sources, this corresponds to about $10^5$ evaluations of the propagation kernel. In the series approach, the eigenvalues and spatial coefficients are computed only once and reused throughout the calculation. By contrast, the integral approach requires a numerical integration for every age--distance combination. For the benchmark setup adopted here, the 500-root series solution requires about $5.2\,\mathrm{s}$, whereas the 160-nodes Gauss-Laguerre integral representation requires about $2400\,\mathrm{s}$. The analytical solution therefore provides a speed-up of nearly three orders of magnitude, while reproducing more stable results.

\bibliography{ms_01}

\end{document}